\PassOptionsToPackage{backref,backrefstyle=two}{biblatex}
\documentclass[biblatex,version=preprint]{iacrcc}
\license{CC-by}
 \usepackage{citeall}
\newcommand{\IACR}{{IACR}}
\usepackage{tikz}
\usepackage{todonotes}

\newcommand{\BibTeX}{{\rmfamily B\kern-.05em%
   \textsc{i\kern-.025em b}\kern-.08em%
   T\kern-.1667em\lower.7ex\hbox{E}\kern-.125emX}}

\newcommand{\hotcrp}{{HotCRP}}

\title[plaintext={Lowering the Cost of Diamond Open Access Journals}]{Lowering the Cost of \\Diamond Open Access Journals}

\addauthor[orcid={0000-0003-1010-8157},
          email={joppe.bos@nxp.com},inst={1}]{Joppe W. Bos}
\addauthor[orcid={0000-0001-7890-5430},
          inst={2},
          email={publishing@digicrime.com}]{Kevin S. McCurley}

\addaffiliation[country={Belgium},ror={0268h4j55}]{NXP Semiconductors}
\addaffiliation[country={USA}]{Self}

\begin{document}

\maketitle
\begin{abstract}
Many scholarly societies face challenges in adapting their publishing
to an open access model where neither authors nor readers pay any
fees. Some have argued that one of the main barriers is the actual
cost of publishing.  The goal of this paper is to show that the actual
costs can be extremely low while still maintaining scholarly quality. We
accomplish this by designing a journal publishing workflow that
minimizes the amount of required human labor.  We recently built a
software system for this and launched a journal using the system, and
we estimate our cost to publish this journal is
approximately \$705 per year, plus \$1 per article and about 10
minutes of volunteer labor per article for the editorial processing. 
We benefited from two factors,
namely the fact that authors in our discipline use \LaTeX\ to prepare
their manuscripts, and we had volunteer labor to develop software and
run the journal. We have made most of this software open source in the
hopes that it can help others.
\end{abstract}

\begin{textabstract}
Many scholarly societies face challenges in adapting their publishing
to an open access model where neither authors nor readers pay any
fees. Some have argued that one of the main barriers is the actual
cost of publishing.  The goal of this paper is to show that the actual
costs can be extremely low while still maintaining scholarly quality. We
accomplish this by building a journal publishing workflow that
minimizes the amount of required human labor.  We recently built a
software system for this and launched a journal using the system, and
we estimate estimate our cost to publish this journal is
approximately \$705 per year, plus \$1 per article and about 10
minutes of volunteer labor per article. We benefited from two factors,
namely the fact that authors in our discipline use LaTeX to prepare
their manuscripts, and we had volunteer labor to develop software and
run the journal. We have made most of this software open source in the
hopes that it can help others.
\end{textabstract}

\section{Introduction}
This paper reports on the experience of launching a diamond open
access journal on behalf of an international scholarly society in mathematics, 
computer science and engineering.
The journal that we
refer to is called \textit{Communications in Cryptology} (CiC),\footnote{See \url{https://cic.iacr.org/}}
and is published by
the International Association for Cryptologic
Research (IACR). As of this date, the journal has published four issues
with 137 articles, and subsequent issues are expected on a quarterly basis.

The work behind this paper started almost ten years ago, when a group
of \IACR\ members first proposed a new peer-reviewed journal with the
goals of rapid publication, open access, and minimal cost.
The society was already
publishing a traditional journal, two journals which operate as a
journal/conference hybrid, and conference proceedings for five annual
conferences. The society also runs a preprint server
at \url{https://eprint.iacr.org} that publishes
about 2100 articles per year. The new journal aims to fill a gap in
the society's publishing portfolio by offering a scalable
publication venue to cope with the rapid growth in our community.

In the last twenty years there has been increasing pressure for
scholarly societies to move toward open access models in which
articles are free for anyone to read. This has resulted in a shift
from the previous subscription-based models to a model in which the
cost of publishing is born by other parties. It has also caused
increased scrutiny on the actual costs associated with publishing.
Some scholarly societies use publishing as a means to raise funds for
other activities. For example, the American Mathematical Society
states that ``The revenue it generates helps support our other
professional activities''~\cite{AMS}.

In contrast, the \IACR\ raises very little revenue through
publishing, and is funded mostly be membership fees and attendance at
conferences. The \IACR\ is a relatively small society, with
approximately 3000 members and annual revenue of approximately \$2M\
per year. \IACR\ pays no staff salaries, although they pay small
amounts to contractors for web software development and accounting
services.  Almost all activities are carried out by volunteers, and
while this is perhaps unusual, it is not unheard of. For example, a
survey~\cite{acls2017} in 2017 showed that among member societies of
the American Council of Learned Societies in humanities and social sciences, 15\% had only a volunteer
part-time executive director and 59\% had annual revenue of
under \$1M.  One study on 1,619 diamond open access
journals~\cite[Findings, p. 8]{diamondstudy2} found that ``60\% of OA diamond journals
depend on volunteers to carry out their work, with 86\% reporting
either a high or medium reliance on them''. The same survey
found that 53\% of diamond open access journals are run on less that 1 FTE
for their operations. This trend toward
scholar-led publishing will be made easier if the need for human labor
is kept to an absolute minimum.

A recent history of publishing at the Royal
Society~\cite{royalsociety} noted that ``...by the turn of the
millennium, staff salaries, overheads and relating costs had replaced
printing and typesetting as the largest component of journal
publishing expenditure at the Royal Society.''  In 2004 it was
reported that peer review and copy editing amounted to ``42\% of the
publishing staff costs''\cite[p. 574]{historyfyfe}.  This highlights
the fact that human labor is probably the most important factor in
controlling costs of publishing.

Over the last 150 years, labor-saving
technologies have revolutionized industries through the industrial
revolution and the information revolution. We believe that this change
is now taking hold in the area of scholarly publishing, and one of our
goals in designing the workflow for our new journal was that it should
use technology to reduce the amount of human labor required. Not only will this keep costs
down, but it can also reduce ``burnout'' by
volunteers~\cite{seismica2023}.
In this paper we share our experience to create a diamond open access
journal which aims to dramatically reduce the amount of human labor
by automating the journal publishing workflow as much as possible. 
This resulted in an total cost of \$705 per year plus \$1 per published paper
for our society. 

\section{Requirements of the new journal}
In planning for the launch of any new journal, the first step should
be to gather the set of requirements and evaluate options.  One
crucial requirement for us was to have the journal follow the diamond open
access model, in which no fees are paid to publish or read the
journal. This necessitates that we keep costs down as much as
possible.  Aside from the obvious requirements of high scholarly
quality, we also wanted to allow ourselves flexibility to innovate in
the highest standards of peer review, copy editing, and indexing.
Requirements may vary from one discipline to another, but some
requirements like metadata collection and facilitating indexing by third parties
are fairly universal.  In this section we will not cover all
requirements that we need to fulfill, but we will focus on a few that
were most influential in our design. As scholarly publishing evolves
in the future, we should anticipate that other requirements may
emerge.

\subsection{Document format}
In many disciplines it is commonplace for authors to prepare their
manuscripts in the Microsoft Word format, but for our discipline this is
almost unheard of. The reason is that our discipline depends heavily on
the ability to produce complex mathematical notation and equations,
and as a result almost everyone uses \LaTeX.
This is not unique to our field, and nearly
90\% of papers\footnote{See: \url{https://arxiv.org/html/2402.08954v1}.} 
uploaded to the popular 
open-access repository of electronic preprints and postprints
\verb+arXiv.org+
were typeset in \LaTeX. This is also not a recent phenomenon, since
already in 1992 over 80\% of papers in journals for the Royal Society were
being typeset in \TeX~\cite[pp. 556]{historyfyfe} either by authors
or staff.

As it turns out, \LaTeX\ is both a blessing and a curse. The \LaTeX\
system is over 40 years old, and it suffers from some archaic
programming practices like a global namespace that lead to conflicts
between packages. It was designed to produce beautifully typeset
articles {\em on paper} with special attention to mathematics, but it
is ill-suited to producing HTML output (see~\cite{arxivhtml}).

On the plus side, it is an extensible markup language that offers the
ability for authors to express their metadata in a structured way. In
a previous paper~\cite{tugboat}, we described how \LaTeX\ can be used
to optimize the capture of metadata from an article, and that became
central to our design to automate the workflow.

Another advantage that \LaTeX\ offers over other formats is that the
output was designed from the start to be beautifully typeset in a
consistent style. This means that the production step of producing the
final published paper often consists merely of compiling the author
source.
Unfortunately, \LaTeX\ is also a programming language and gives a
great deal of freedom to authors to deviate from the style. Our system uses
checks to try to ensure major deviations won't happen, but it requires some
cooperation from authors.

Underlying all of this is the assumption that the final product is
something like PDF that emulates paper. That has several drawbacks
such as the inability to reflow text to render on small screen
sizes. It is also difficult to produce an output that can be
understood by assistive technology for readers with sight impairment.
The \LaTeX\ team is engaged in a multi-year project to make PDFs produced
by \LaTeX\ more accessible, and we expect to see improvement in this.
That still wouldn't solve the problem of reading on small screens, 
for which HTML would be a better format than PDF. 
There has been quite some progress in this direction~\cite{arxivhtml}
with positive outcomes but the project remains experimental.

One of the strengths of the \LaTeX\ system is that authors are able to
typeset their articles themselves. They validate
their own work when then they submit it, but that is partly because
they mostly use a standard \LaTeX\ distribution.  The journal is
essentially placing the responsibility for typesetting in the hands of
the authors, but within a system where they can easily verify their
work. We are reluctant to place an additional burden on authors to
validate their HTML output, so in our first launch we just produce an
HTML version of the metadata, abstract, and references. We hope to
backfill full HTML for existing papers in the future. 

\subsection{An ideal document format}
The difficulty of producing accessible output that displays well on
small screens highlights the fact that \LaTeX\ is not an optimal
format.  The same can be said about the Microsoft Word \texttt{docx} format
that is commonly used by authors in non-STEM fields, and in fact
the situation is probably even worse there because the \texttt{docx}
format has minimal support for structured metadata.

In order to produce both HTML and PDF, many publishers invest considerable human
effort to convert the original author format into another format that
is more suited to long term preservation and conversion to other
formats. Perhaps the most popular of these is the Journal Article Tag
Suite (JATS~XML)~\cite{jats}.  One of the primary motivations for this
format is to capture the structure of scientific documents along with the
metadata. 

One problem with JATS is that almost no authors have been writing in
this format, and conversion of \LaTeX\ or \texttt{docx} into something like
JATS or HTML usually requires human intervention and is likely to lose
the visual nuances that authors exploit to express their ideas.
This {\em impedance mismatch} between what authors use and what
publishers want is part of the problem of reducing the cost of
publishing. As long as we base a publishing system on \LaTeX, we have
to accept the inherent limitations that it imposes. We believe that
if \LaTeX\ is to survive in the modern age of publishing, then it
should be less tied to paper and PDF, and more reflective of document
structure.  This will also require authors to think differently about
their articles when they produce them. 

\subsection{Requirements for peer review}\label{requirements}
Like most parts of
computer science~\cite{fortnow2009viewpoint}, most of \IACR's publishing has been in conference
proceedings. As a result,
most peer review in the field is carried out in a somewhat different
model than in other disciplines, using a public committee of
reviewers, with at least three reviewers assigned to read each
paper. In some cases the committee may request a review from a
reviewer who is not on the committee. This results in a
reviewing process where authors are hidden from
reviewers, but most reviewers are known to have been selected from a
known committee. 
All of the \IACR\ conferences and two of \IACR's
journals use the \hotcrp\footnote{See \url{https://github.com/kohler/hotcrp}} open
source system for their peer review.  The \hotcrp\ system was designed
for conferences, in which there is a fixed deadline for all
submissions to be made, and reviews and decisions are made according
to a pre-approved schedule.  This fits the model for two of \IACR's
journals, because submissions are made on a schedule four times a
year, and rapid decisions are made on a predictable schedule for each issue.

The entire process of peer review and publishing is under intense
pressure in recent years, in part due to rapid growth in the number of
publications~\cite{10.1162/qss_a_00327}.  For example, the {NeurIPS}
conference in machine learning received 13,300 submissions in 2023,
resulting in the acceptance of 3,540 papers. At this scale, it becomes
very difficult to identify and recruit reviewers, detect conflicts of interest and
discrimination, and provide a uniform level of consistency across reviewers.
The \IACR\ has not grown at that scale, but there is clearly an increasing need for
tools to optimize the publishing process.

Another problem with a traditional journal blind peer review system is that editors are
increasingly having a hard time finding reviewers because there is no
clear incentive for reviewers to do the work. One advantage of the
conference-style reviewing is that reviewers are incentivized to
perform this important reviewing task because they get public
recognition for having done so.  Another approach that seeks to
address this is the open peer
review system~\url{https://openreview.net} that was
described in~\cite{openreview}.  In this model, reviews are published
along with the papers. This can help to give credit to reviewers and
help readers to understand the context of the paper through
independent assessments. This is sometimes even done for papers that
are rejected. There are numerous other suggestions that have emerged
to improve upon peer
review~\cite{bengio,endofanerror,IsPeerReviewABadIdea2021,payreviewers,10.1002/leap.1544}.
For this reason, we believe that flexibility in planning of peer
review is crucial.

\section{Existing platforms}
Our model of peer review played a prominent role in our planning to
launch a new journal, because we wanted the flexibility to modify and
improve the peer review system. When we talked to commercial
publishers about this, they seemed to be unprepared to accommodate
because it did not fit the traditional model of peer review that their
software had implemented.  Moreover, the diamond open access model
is inconsistent with the profit goals of commercial publishers.

We already run several systems based on open source software, so the
natural next step was to evaluate existing open source platforms for
journal publishing.  The two most obvious candidates were Open Journal
Systems (OJS)~\cite{ojs} from the Public Knowledge Project and the
Janeway system~\cite{Eve-2018} from the Open Library of Humanities.
At the time that we started our own project, IACR was already
publishing two diamond open access journals on OJS through a
partnership with Ruhr University Bochum.  This is perhaps an
exaggeration, because we were not using the submission or production
parts of OJS, but were instead using a slightly modified version
of \hotcrp\ for the submission and reviewing, and we had volunteers
doing the copy and production editing manually outside of any existing
system. The only part of OJS that we were using is the ``quickSubmit''
plugin, which turns out to be fairly labor intensive because each
paper must have the metadata entered by hand into a form.
Perhaps most seriously, OJS offers essentially no direct support for
the \LaTeX\ platform, and all copy editing is assumed to be done
outside of the platform itself. This makes it clumsy to communicate
with authors about changes made to their papers and track responses
from authors about individual changes.

The only alternative to OJS that we found at the time was
Janeway,\footnote{\url{https://janeway.systems}} but after inspecting
it, we concluded that it suffered from some of the
same problems that OJS had. We couldn't use their peer review system,
but luckily Janeway has better support for importing articles than OJS
does. Janeway also has no direct support for \LaTeX, which was also
observed in~\cite{jcls_3627}.
As a result, all copyediting and production tasks must be handled
outside of the system, with the results uploaded. Once again this is a
fairly labor-intensive task, and fails to take advantage of the fact
that
\LaTeX\ is already designed to produce high-quality typeset material.

\section{Components of a publishing workflow}\label{components}
After evaluating the different options, we decided
that there is no open source software for running a journal that
supported our requirements, and it appeared to be very complicated to
customize existing platforms. Because of this,
we implemented a new open-source software system for the publishing
workflow.

As we set about building a software system to support our publishing workflow,
we wanted to limit the task to just what we missed. We were already using
a slightly modified version of \hotcrp\ for submission and reviewing of papers for two other journals,
but we also wanted to leave ourselves freedom to substitute another
system like \verb+openreview.net+. There is good reason to believe
that we may see changes in other aspects of academic publishing,
including post-publication peer review, more flexible and interactive
media formats like HTML, versioning, and coupling to experimental
results, software, and data. This is daunting for a publisher, because
it greatly complicates the planning of any software system to
accommodate future change.  Moreover, publishing workflows can be
quite complicated, with multiple parties such as authors, editors,
reviewers, sub-reviewers, copy editors, production editors, and system
administrators. They can also involve third parties such as indexing
agencies, plagiarism services, preservation organizations, and DOI
issuers.

Assuming that we at least adhere to a pre-publication peer review
system, it simplifies the design process to think of the publishing
workflow as being broken into three phases (see
Figure~\ref{phases}). We were already using something like this for
two of IACR's journals, where the peer review took place in \hotcrp\
and the accepted papers were transferred to {OJS} for indexing and
hosting. By breaking the publishing process into these separable
components, we maintain some flexibility in changing the parts that we
need to.

The thing that characterizes these phases is the fact that there is
relatively little need to integrate across all of these. For example,
the review and approval phase does not need to know about copy editor
IDs. The output from one phase remains frozen thereafter, and can be
processed independently by the next phase.  The output from the peer
review phase is essentially just a list of authorizations for authors
to submit their final versions, along with minimal data about document
ID and submission and acceptance dates. The output from the copy
editing and production phase is a list of finalized publications along
with their metadata. In our implementation, the submission and
reviewing phase is handled by \hotcrp, and the copy editing and
production phase is handled by a separate server whose code is
described in section~\ref{codestructure} and available as open
source~\cite{latexsubmit}. The indexing and web hosting is handled by
yet a third server that we plan to make open source, but this
part is relatively simple compared to the other two components.

\usetikzlibrary {shapes.symbols, intersections, arrows.meta}
\begin{figure}[t]
\centering
\begin{tikzpicture}
 \node [fill=cyan!20, cloud, cloud puffs=11.7, draw, minimum width=3cm, minimum height=2cm, name=cloud1]
   at (1, 1) {};
 \node[align=center,scale=.8] at (1,1) {Submission\\ and\\Reviewing};
 \node [fill=cyan!20, cloud, cloud puffs=10.7, draw, minimum width=3cm, minimum height=2cm, name=cloud2]
   at (6, 1) {};
 \node[align=center,scale=.8] at (6,1) {Copy editing\\ and\\production};
 \node [fill=cyan!20, cloud, cloud puffs=11.2, draw, minimum width=3cm, minimum height=2cm, name=cloud3]
   at (11, 1) {};
 \node[align=center,scale=.8] at (11,1) {Indexing and\\ web hosting};
 \draw[-Stealth] (cloud1) -- (cloud2);
 \draw[-Stealth] (cloud2) -- (cloud3);
 \node[align=center, scale=.75] at (3.5,1.6) {accepted\\ papers\\ for issue};
 \node[align=center, scale=.75] at (8.5,1.6) {final\\ papers \& \\ metadata};
\end{tikzpicture}
\caption{A publishing workflow can naturally be broken into three phases, with simple flow of data from one
phase to the next.}
\label{phases}
\end{figure}

The modularization of a publishing workflow allows us to mix and match
different components to perform the phases, and all we have to do is
define the data that passes between them.  These phases are implemented as independent
web servers, though they could also be implemented on a single machine.
When a paper is accepted in the review
phase, the author is directed at an authenticated URL in the copy
editing and production phase where they can upload their paper. When
papers for an issue are approved from copy editing, the copy editor causes
the paper to be recompiled into the final version. When an issue is complete,
the editor exports a bundle to the indexing and hosting phase that contains
the final PDF versions of the papers, along with the metadata for each
paper and some minimal metadata to describe the issue.  This exported
bundle becomes the input to the indexing and web hosting phase. This phase
is relatively simple, because it registers the DOIs for the papers, imports
the metadata for the papers into the content management system, and exports
the issue.

Note that after a paper is accepted for publication, there are some
metadata elements that may change. These include things like the
title, email addresses, affiliations, and even author names. In order
to properly propagate this metadata forward between the phases, we
need to make sure that it gets updated. We solve this problem by
making sure that the metadata is specified in only one place, namely
the source documents supplied by the corresponding author. It is the
job of our \LaTeX\ class to process this metadata when the final
version is uploaded, and there is no need to propagate it from the
review and approval system.

\subsection{Typesetting and copy editing}\label{codestructure}
Before we describe our approach to copy editing, we should mention
that there is some disagreement about the value of copy editing in
scientific literature. Nobody will disagree that a better written
paper results in a more impactful piece of work, but a major question
is: who should bear the responsibility for better writing? Having
someone look at a paper to correct grammatical errors will cost money,
and part of our goal is to minimize costs. Some societies
like ACM\footnote{See \url{https://www.acm.org/publications/pacm/pacm-guidelines}}
and AMS\footnote{See \url{https://www.ams.org/arc/journals/index.html}.} offer
access for authors to discounted copy editing services through a third
party commercial service. Authors are responsible for paying for the
service.

Over the years, some tasks that were traditionally done by copy
editors have been automated. The most obvious one is spelling
correction. A more sophisticated approach is used in the
{LTeX} system,\footnote{See \url{https://valentjn.github.io/ltex/}} which
uses Language Tool\footnote{See \url{https://languagetool.org}} to
identify various grammatical errors. This can be used to recognize
relatively simple things like incorrect verb tense, improper
capitalization, and improper use of punctuation. We have experimented
with this, but found that it was often fooled by the complex notation
of a mathematics article. The result was that it produced too many
false positives that could just annoy the author(s).

The last few years have seen enormous advances in use of large
language models for generative artificial intelligence (GAI). Some
disciplines regard this with alarm, since it can generate text that is
essentially indistinguishable from text that is produced by a
human. Essentially all journals have been scrambling to craft a policy
about the use of these technologies, but we see large potential
benefit in the future through application to automated copy
editing. If this comes to fruition, it will be yet another example of
labor-saving technology being applied to scientific publishing.
We look forward to experimenting more with these techniques in the future,
but for now we are leaning heavily on the compilation process of \LaTeX\,
and providing automated feedback to authors for them to fix commonly
occurring problems.

When an author's paper is accepted in the submission and review phase, they
are presented with a personalized URL that they can use to upload the final
version of their paper. We actually refer to that version as the ``candidate version'',
because it represents what the author thinks is final but has not yet gone
through copy editing or production. The upload consists a zip file containing
a file \verb+main.tex+ along with any other \TeX\ files that are required
to compile their paper. We also require the author to upload \BibTeX\ file(s),
because we use a structured form of the bibliography as metadata that
is reported to indexing agencies (see Section~\ref{indexing}).
We also parse the \BibTeX\ file to check for missing fields and we produce
an HTML version of the references.

\begin{figure}[t]
\centering
\includegraphics[width=\textwidth]{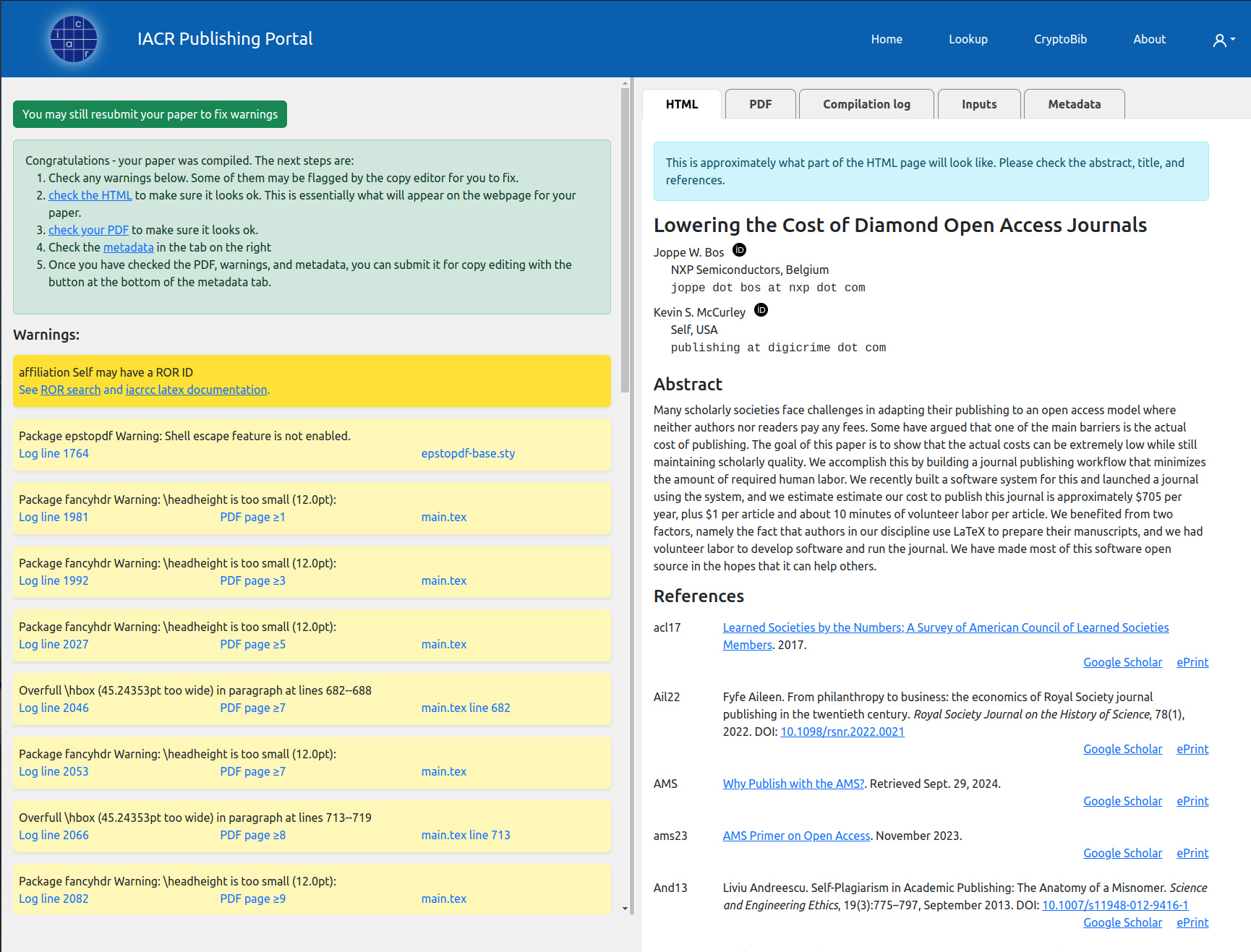}
\caption{View shown to the author with structured feedback from the \LaTeX\ and \BibTeX\ logs.
The left column shows a sequence of items that were flagged during the
compilation and extracted from the \LaTeX\ and \BibTeX\ logs. The
right column has tabs to view the HTML, PDF, logs, etc. We rely
heavily on authors to fix these problems, but we provide them with a
much better way to spot the problems.}
\label{candidate}
\end{figure}

When the author uploads their candidate version, the server
immediately compiles it in the cloud in a sandbox environment and reports back to the author
anything that needs to be changed. The log files from the compilation
would ordinarily contain this information, but they are so verbose and
hard to understand that authors mostly ignore them.  In order to
overcome this, we wrote a parser for the output logs from the \LaTeX\
and \BibTeX\ compilations. This extracts only the serious errors and
warnings and presents them to the author in a more readable structured
format with pointers to where the problem occurs in the source files
and the PDF. A view of the author's feedback can be seen in Figure~\ref{candidate}.
Our \LaTeX\ class is designed to extract all metadata at compilation time,
which allows us to automatically generate the HTML rendering. Part of the
review process involves having the author review this HTML rendering.

We have found that one thing authors often overlook is the inclusion
of DOIs for journal article references, so we flag those. We also have
a user tool for the author to automatically find these DOIs via
search. The author should fix any errors or serious warnings, and
keep uploading their source files until they are satisfied with it.  At
that point the paper moves to the copy editing phase and is recompiled
with line numbers in it.

Once the author has responded to all of the automated feedback and
forwarded their document for copy editing, we employ a minimalistic
technique of performing only a cursory scan of the document by a human
copy editor. The copy editor looks at both the things flagged by
the \LaTeX\ compilation but also performs a few simple checks to
confirm that the paper conforms to the style of the journal. This may
of course allow some papers with low grammatical quality to be
published, but in our view this reflects more on the effort put in by
the author(s) than the journal. This is not a new problem, and
to quote Charles Babbage~\cite{babbage} from 1830,
\begin{quote}
``With regard to the published volumes of their Transactions, it may be
remarked, that if members were in the habit of communicating their papers
to the Society in a more finished state, it would be attended with several advantages;...''.
\end{quote}

We do {\em not} assume that the copy editor is skilled in \LaTeX,
and we do {\em not} expect the copy editor to edit the author's source files.
In practice our copy editors are volunteers from the field, so they have at
least some basic understanding of what likely happened to cause a problem
in the paper.
We merely send items back to the author listing the page number, line number,
source file, and line in the source file where the problem should be fixed.
It is the responsibility of the author to fix things, and we have found them to
be mostly cooperative in this process.

\begin{figure}[t]
\centering
\includegraphics[width=\textwidth]{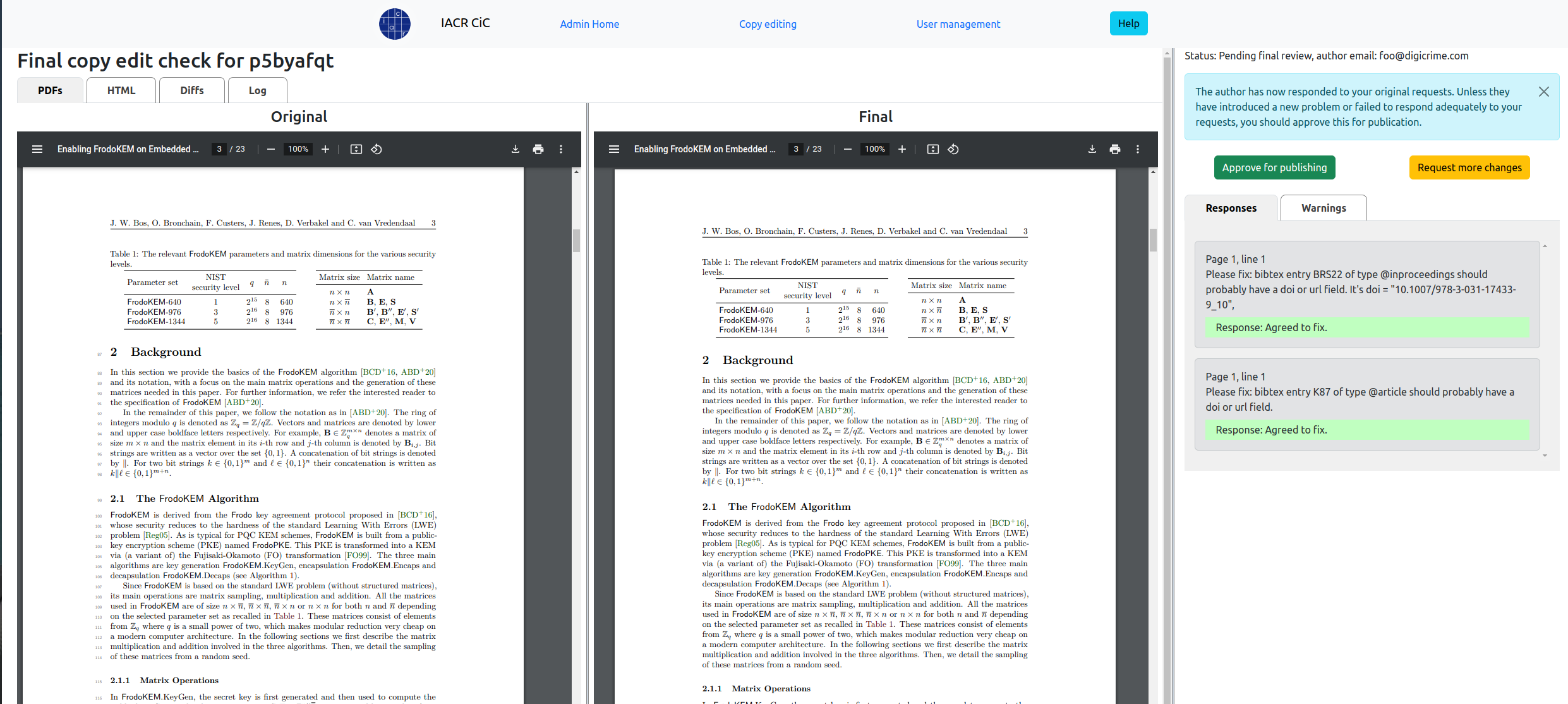}
\caption{View shown to the copy editor after the author has sent responses to all items flagged for changes.
The copy editor can view both the before and after of the PDFs as well as the differences in the \LaTeX\
sources from the two versions. In this case the author agreed to both changes that were requested, so
the copy editor can check the result.}
\label{copyedit}
\end{figure}

The author is required to respond to each item that is sent to them by
the copy editor, either to say they will fix it, they will not fix it
(and why), or to request further clarification. This can theoretically
generate another round of copy editing where the editor sends it back
to the author to fix.  The view in Figure~\ref{copyedit} shows
what the copy editor is shown after the author has responded to all of
the items they were sent during the copy editing phase.

We have allocated a total of one month to complete this process from
the time that the author is notified of acceptance and is told where
to upload their ``candidate'' version.  If the authors do not meet
this deadline, then their paper will be deferred to a later
issue. This is quite different from some commercial publishers, where
they spend considerable time reformatting and editing the document,
and then often give the author only a few days to respond to changes
in their final version. In our experience the copy editor only needs
to spend 5-10 minutes per article. The copy editor doesn't need to
download anything and they don't need to directly edit the author documents.
In practice we have observed that authors often create extremely complicated \LaTeX\
constructions, so it is better left to the author to fix their problems anyway.

It should be noted that our estimate for the amount of time for copy
editing and production is dramatically lower than what was described
in~\cite{f1000}, where they estimated that all editing duties would
require a total of 7.5 person-hours per article. In~\cite{olhblog}
they estimated that three people could perhaps handle 1000 articles
per year. That comes out to about 6 hours per article assuming a 40
hour workweek spread over 50 weeks. We achieve such a large advantage
because we use \LaTeX\ for typesetting, and our system automatically
highlights many potential copy editing issues.

A demonstration video of this system is available
at \url{https://youtu.be/Ki8Qcai9klg}, and a test instance is currently
available online at \url{https://publishtest.iacr.org/}.

\subsection{Indexing and web hosting}\label{indexing}
From the beginning of this project we were keenly aware of the need to
produce not only a finished paper to read, but also of the need for
metadata to facilitate indexing and discovery.

In the process of our workflow, we rely upon our \LaTeX\ class to
extract significant metadata from the author's source during
compilation~\cite{tugboat}. This includes the title, subtitle, author
names, author affiliations, funding information, a text abstract, and
all bibliographic references. Just as articles are now identified by
DOIs, we also request ORCID\footnote{See \url{https://orcid.org/}} and
ROR\footnote{See \url{https://ror.org/}} identifiers
for affiliations and funding agencies.
We report these to \verb+crossref.org+ in
the course of registering the DOI for the paper, and we also expose a
significant amount through an OAI-PMH interface.\footnote{See \url{https://www.openarchives.org/}.}

\section{The cost structure of scholarly publishing}
The publisher Stanley Unwin~\cite{unwin26} has often been quoted as saying ``A publisher's first duty to
his authors is to remain solvent''.  It is therefore imperative that we analyze the cost
structure for publishing evaluate whether it is sustainable.
While there have been several studies on what publishers {\em charge}
for open access publishing, there is relatively known about what the
actual costs are. One notable study~\cite{f1000} estimated that the
actual cost of publishing open access is probably under \$400 per article,
and one estimate from 2010~\cite{edgar2010survey} estimated the cost at \$188
per paper. The purpose of this
article is to show that the actual costs for publishing can be
considerably lower than all of these estimates, even without
sacrificing quality.

It was observed in ~\cite{f1000} and~\cite{olhblog} that the cost structure
for a publishing operation has both fixed costs as well as some per-article
costs. As a result, a per-article cost may not be
the correct metric to measure the actual costs, but they are useful
to compare against APCs charged by journals. In all cases a publisher can achieve an
economy of scale from running a larger operation that publishes more articles.

One way to estimate the cost of publishing is to examine the process used by arXiv.
There have been several studies that tried to estimate their cost per paper.
By examining the annual report from
arXiv~\cite{arxivbudget}, it appears that their revenue in 2023 was \$3.36M and they
published a total of 208,493 articles. This works out to a cost of \$16.11 per paper, but
it assumes that all effort is directed toward publishing (no outreach, fundraising, etc).
Another study~\cite{10.1002/leap.1219} estimates their cost per paper as \$9.70 in 2018,
but arXiv's budget has increased substantially since then due to some grants they received.
An older study~\cite{VanNoorden2013} estimated their cost per paper at approximately
\$10 per paper in 2010. It should be pointed out that arXiv does not perform all of the
tasks that a full journal publisher would perform. They use simple moderation instead of
peer review, and they perform no copy editing. They do at least assign DOIs and they
compile submissions in \LaTeX.

A recent study commissioned by the European Commission for Open
Research Europe~\cite{ore2023} was quite a bit more pessimistic. There
are many assumptions built into their model, including the
establishment of an independent organization with a publication rate
of 2000 articles per year and a paid staff of 7-10 full time
employees. In \cite[Section~6.5]{ore2023} they estimate a per-article
cost of €2,115 per article in
2030. Their analysis points out that the volume of publications is a
crucial factor in determining the per-article cost of publishing.
In fact, they estimate the article production cost at €650 per article~\cite[Table 6]{ore2023},
so the fixed costs are dominant.
This is also reflected in the analysis of what it costs for arXiv to
publish an article, because they have very high volume. In our opinion
the report simply highlights once again the need to focus on
technology that will minimize the need for human labor.

\subsection{Breakdown of our cost structure}
In estimating our cost structure, we need to consider startup costs,
yearly maintenance costs, and per-paper costs. Our startup costs
consisted merely of the cost to purchase an ISSN identifer for the
journal, namely €53. Since we wish to assign DOIs to our
articles, we needed to join an organization like Crossref.  For
organizations like IACR that make essentially nothing from publishing,
this is currently \$275 per year.  Then we need to have a cloud
service to run our machines. We use Digital Ocean, and our cost for
running the machine is \$180 per year (including backups). Based on
our experience in running a preprint server, we estimate that we could
easily run at least 10 journals on this single machine.  We already had a web
service for the society, so there was no additional cost to register
an internet domain.

We run our own open-access versions of \hotcrp\ for the review and
submission phase.  This is currently running on an additional machine
at a cost of \$14/month (including backups), but we could easily run
it on the same machine.  In our case we run it on a separate machine
because we also use it for reviewing of eight conferences per year plus
two other journals. Even with this load it averages only 3\% CPU usage.
For this reason we omit the additional cost of
running \hotcrp.

The creator of \hotcrp\ runs a hosted service
for those who don't wish to run their own copies of HotCRP. He
currently quotes a price of
\$7.50 per submission. In our case, the journal an
acceptance rate has been about 38\%, so we estimate the cost per
published article as approximately \$20 to contract out the submission
and review component. For those who prefer to use a more traditional
peer review system, the Scholastica service currently offers a
peer review system priced at \$10/submission plus \$350 per
year.\footnote{See \url{https://scholasticahq.com/pricing/}} We expect
the costs would be similar for either a self-hosted OJS or using a
hosted service.

One requirement placed on all electronic journals is to have a long
term preservation plan so that the content of the journal will always
be available in the case the society goes away or stops hosting the
journal.  There are several such services that a journal can subscribe
to, including the Portico service. Their annual fee for a small
publisher like IACR is \$250 per year. One alternative for a diamond
open access journal is to depend upon the Internet Archive to archive
everything for free.  There is however no quality of service guarantee
unless the publisher purchases such a service.

Next, we have costs per article. Assignment of a DOI from Crossref
costs \$1 per article. Some publishers subscribe to an additional
service for similarity checking to detect plagiarism, but we rely upon
our broad editorial committee to detect these. We also believe that
plagiarism in mathematics is fairly rare, though a few cases have
been identified.\footnote{See \url{https://retractionwatch.com/category/by-subject/physical-sciences-retractions/math-retractions/} and
\url{https://www.acm.org/binaries/content/assets/about/annual-reports/pubs-annual-report-fy23.pdf}}
Even if we signed up for this service, it would only increase
our crossref fees by 20\% per year and \$0.75 per article.

So far this amounts to \$705 per year fixed costs if we neglect the
plagiarism service, plus \$1 per article. Our current expectation is
to publish approximately 200 articles per year, which works out
to \$905. If we were to publish 1000 articles, the cost would
be \$1705 per year.

\subsection{Cost of human labor}
All of this neglects the most important part, namely human labor.
Like most scholar-run journals, the administrative management of CiC
is done by volunteers, but that may not be
feasible for journals in other disciplines because of the skills required.

The biggest amount of time involves the effort to perform reviewing of
papers.  It is almost impossible to estimate the total amount of time
spent by humans to referee articles in mathematics and computer
science that are submitted to a journal. One estimate says at least
hours~\cite{refereeing} and perhaps days. A survey of academics across
multiple disciplines~\cite{ware} found that the median time spent per
article was 5 hours and the mean was 9 hours.  Another recent
article~\cite{LeBlanc2023} estimates the time at four hours. They also
estimate that the total cost of labor for performing peer review is
approximately
\$6~billion. This labor is typically unpaid, since academics and researchers perform
this task as part of their normal duties for their employers but we
feel compelled to mention it because it's by far the largest amount of
human labor. The fact that our journal is diamond open access has
helped to recruit reviewers and editors which are willing to 
dedicate their valuable time.

In addition to the time spent reviewing, the editors have to devote a
considerable amount of time in their role of selecting editorial board
members, assigning reviewers, overseeing the decisions resulting from
reviews, and performing ``desk rejects'' on articles.  Some commercial
publishers pay a salary to editors, but most editors of journals run
by non-profits perform their services without pay as part of their
normal duties.\footnote{There are exceptions to this. The Editor in Chief of the
``non-profit'' AAAS journal \textit{Science} was paid \$654,263 in
2023. See \url{https://projects.propublica.org/nonprofits/organizations/530196568/202403209349318715/full}.}
Faculty members at research universities are often expected to
divide their time into 40\% research, 40\% teaching, and 20\% for
service. Editing a journal is usually counted under the service part.

Aside from the labor for performing editing and reviewing, the rest of
the labor is dedicated to IT tasks.
In our case we are lucky to be able to draw upon a pool of
volunteers in computer science who have significant skills in
information technology. In some disciplines this kind of expertise is
less common, but university IT departments and libraries often have
this kind of expertise. The specific kind of skills we needed included
software development and linux system administration, and will be
described in the next two sections. Any organization that runs a web
server is likely to have these kinds of skills.

\subsubsection{Software development}\label{softwaredev}
We had several pieces of software to develop. First, we developed a \LaTeX\ class file
that accomplished our goals for metadata handling described in~\cite{tugboat}.
The basic techniques to facilitate metadata extraction would be easy to
adapt to a different journal style.
Next, we had to develop software for the three phases described
in section~\ref{components}. For the reviewing phase, we made fairly
minor modifications to HotCRP to facilitate the passing of data from HotCRP to the copy
editing and production phase. That was only a few hours of work. Next we had to design and build
the copy editing and production phase, which is the most complicated part.  We estimate
that this took approximately 18 person-months for this, resulting in a system
with about 16000 lines of python code. Finally, we had to
develop our system for indexing and hosting. That took approximately three months of work
to produce about 6000 lines of code.
The code used for the first two phases are now open source so others may be able to adapt it to their
needs in significantly less time. We also plan to release the code for
the indexing and hosting phase as open source.

\subsubsection{Systems administration}
We had to configure a linux virtual machine with a cloud provider in
order to run the web servers for the phases of our workflow. For many
IT workers, this is now a fairly routine task that is now easy to
accomplish if someone is knowledgeable about linux. Theoretically the
software could also be run on a windows server, though it would be more
complicated to accomplish.

Once the server has been set up, we needed to configure and install
the software from section~\ref{softwaredev}. As an ongoing task,
servers will require occasional updates, but that amounts to maybe an
hour a month. We also need to update and test software packages
required by our applications, but that is also a minor amount of time.
We estimate that this labor might cost \$200 per month from a contract
system administrator who is paid hourly, but it should be a minor task
for any in-house IT person to manage if they are already managing a
web server. 

\subsubsection{Economy of scale}
Several studies~\cite{diamondstudy,Wise-2019,ore2023} have identified
the importance of sharing resources between small society publishers,
because the cost of labor or IT services can be the dominant factor.
This is particularly important in the development of software.  There
are more than 25,000 journals being run on the OJS software
platform~\cite{10.1162/qss_a_00228}, and one
study~\cite{diamondstudy2} found that 60\% of their respondents
mentioned running their journals on the OJS platform. If they had all
been required to invest in developing their own platform, then diamond
open access would have been severely handicapped. Moreover, a single
installation of OJS can be used to operate multiple journals, so the
effort for installing and maintaining an instance of OJS can be
amortized over multiple journals. Similar claims can be made for the
Janeway platform and the platform we developed. There are several
different services that will install and operate OJS and Janeway for
journals at relatively low cost (e.g., an OJS installation for
under \$1000/year).

In our case we only sought to publish one journal with a throughput of
200 articles per year, so the amount of volunteer effort that is required
is fairly moderate. We expect that the same infrastructure could easily
support a throughput of 2,000 articles per year and perhaps even up to 10,000 articles
per year. Unfortunately at such a small scale it can be difficult
to identify a source for the small amount of IT support required to
set up and maintain the system. As we mentioned, we expect it require only
a few hours a month from a competent person, but that is not enough work to
justify hiring someone. This is where services to host popular platforms
like OJS and Janeway for many journals start to make sense.

\subsection{Summary of our costs}
Taking all of our costs into account, we estimate that our total cost to operate
a diamond open access journal amounts to under \$1000 per year to publish 200
articles. We are aware that we rely heavily on volunteer labor to handle some
tasks, but we are not the only ones who have been able to overcome this~\cite{seismica2023}.
Experience has shown that if a group of scholars or scholarly society wants
to operate a diamond open access
journal for low cost, then it is now completely feasible to do so.
While there are real costs associated with
running a diamond open access system, the costs can be quite low.

\section{Future directions}\label{future}
While we feel that we've accomplished the original goal that we set out to solve,
there are a number of areas for future improvement. For example, our system for 
copy editing and production is currently dependent upon use of a single
\LaTeX\ class for our journal. In order to make this more accessible to
other publishers, we have separated out the metadata capture parts as
a separate \LaTeX\ package called \texttt{metacapture.sty}.\footnote{See
\url{https://github.com/IACR/latex/tree/main/metacapture}} This should allow other journals to
apply their own styling. We also plan to define a better API for our
system to be used as a plugin to other systems such as OJS,
Janeway, and openreview.net.

It remains to be seen whether this approach can reasonably be applied
to disciplines in which authors write in the Microsoft Word \texttt{docx}
format~\cite{ooxml}.  One problem with this format is that it does not
define suitably detailed format for metadata of journal articles, in
spite of the 5000 pages in the specification. This makes it difficult
to convert directly from \texttt{docx} to JATS or even well-structured PDF
because the metadata needs to be constructed from a source that is
external from the author's original document.
Given the current state of the art, we suspect that human
intervention will always be required in this conversion, which
conflicts with our goals of eliminating the need for human labor of
editors.

As we mentioned in section~\ref{future}, it would be very desirable to
have high-quality HTML output produced automatically from the
author's \LaTeX\ source.  Several groups have tackled this, but it
remains an open problem. If this is accepted as a requirement for a
journal, then they will either have to place restrictions on the kind
of \LaTeX\ that is used by authors, or they will have to devote human
labor to fixing the output from the conversion tools that exist.

Some journals may object to the lack of human-driven copy editing that we have
adopted. We are hopeful that the use of LLMs for copy editing will answer
these concerns.

\section{Conclusions}
In analyzing the workflow for running an academic journal, we have
identified several places where the flow can be improved with the
assistance of better technology. We were pleasantly surprised to find
that authors were active partners in making sure that the process went smoothly,
and we are confident that the process can be sustained with mostly volunteer effort.
The actual financial costs turn out to be quite low, and it provides hope for
the future of diamond open access publishing. We hope that open standards can
help to promote this activity in the future.

\printbibliography
\end{document}